\documentclass[9pt,
reprint,
superscriptaddress,
showkeys,
amsmath,amssymb,prl
]{revtex4-2}
\usepackage{xcolor}
\usepackage{xfrac}
\usepackage{graphicx}
\usepackage{dcolumn}
\usepackage{bm}
\usepackage{amsmath}
\renewcommand{\eqref}[1]{Eq.~(\ref{#1})}

\begin{document}

\title{Inverse reconstruction of dissipative Kerr soliton interactions}%

\author{Carlo Silvestri}
\email{carlo.silvestri@sydney.edu.au}
\affiliation{Institute of Photonics and Optical Science (IPOS), School of Physics, The University of Sydney, NSW 2006, Australia}
\affiliation{ARC Centre of Excellence for Optical Microcombs for Breakthrough Science (COMBS), School of Physics, The University of Sydney, NSW 2006, Australia}

\author{Panayotis~G.\ Kevrekidis}
\affiliation{Department of Mathematics and Statistics, University of Massachusetts,Amherst, MA 01003-4515, USA}

\author{St\'ephane Coen}
\affiliation{Department of Physics, University of Auckland, Auckland 1010, New Zealand}
\affiliation{The Dodd-Walls Centre for Photonic and Quantum Technologies, Dunedin, New Zealand}

\author{C. Martijn de Sterke}
\affiliation{Institute of Photonics and Optical Science (IPOS), School of Physics, The University of Sydney, NSW 2006, Australia}
\affiliation{ARC Centre of Excellence for Optical Microcombs for Breakthrough Science (COMBS), School of Physics, The University of Sydney, NSW 2006, Australia}

\author{Antoine F. J. Runge}
\affiliation{Institute of Photonics and Optical Science (IPOS), School of Physics, The University of Sydney, NSW 2006, Australia}
\affiliation{ARC Centre of Excellence for Optical Microcombs for Breakthrough Science (COMBS), School of Physics, The University of Sydney, NSW 2006, Australia}
\begin{abstract}
We reconstruct the pairwise interaction potential between dissipative Kerr solitons from the linear stability spectrum of a perfect soliton crystal solution of the Lugiato--Lefever equation. The solitons form an overdamped lattice whose positional eigenvalues encode the pairwise-force derivative, mirroring the extraction of microscopic interactions from phonon or relaxation spectra in condensed-matter systems. The method extends beyond quadratic dispersion and assumes no functional form for the interaction. Our theory predicts a novel state---the soft soliton crystal---which has vanishing stiffness and a diverging positional relaxation time, and also explains experimentally observed soliton steps as arising from a sign reversal of the pairwise-force derivative.
\end{abstract}
\maketitle
\textit{Introduction}---Collective excitations provide a powerful window into the microscopic interactions underlying ordered many-body systems~\cite{BornHuang1954}. This principle is widely exploited in condensed-matter physics: in conventional crystals, for example, interparticle force constants can be inferred from phonon spectra~\cite{Mohr2007,Flensburg}. In strongly damped systems, such as colloidal crystals, the corresponding information resides in relaxation rate spectra rather than oscillation frequencies~\cite{Keim2004,Baumgartl2008}. At the heart of this approach lies a connection between collective dynamics and microscopic interactions: rather than probing individual interactions directly, they are inferred from the response of the assembled system.


Nonlinear optics is a natural setting to which to extend this paradigm. Indeed, localized pulses in passive nonlinear resonators, self-organize through their mutual interactions into ordered configurations known as soliton crystals~\cite{Cole_2017,Karpov_2019}, providing an optical analog of an interacting particle lattice. These pulses, \textit{dissipative Kerr solitons} (DKSs), are sustained by the double balance between cavity dispersion and Kerr nonlinearity, and between external driving and cavity losses~\cite{Kippenberg_2018,Pasquazi_2018}. In frequency, they manifest as optical frequency combs, phase-coherent sets of equally spaced spectral lines~\cite{Kippenberg_2018,Pasquazi_2018,Hansch}. The physics of DKSs is accurately described by the Lugiato--Lefever equation (LLE), a one-dimensional nonlinear Schrödinger equation incorporating coherent driving, linear loss, and cavity detuning~\cite{LLE_1987,Coen_2013}. 

Soliton crystals arise within the standard LLE, where pulses adopt an equidistant configuration, forming a perfect soliton crystal (PSC)~\cite{Gomila2022}. The formation of soliton crystals can also be promoted by additional mechanisms~\cite{Cole_2017,Karpov_2019,Lu_2021,Wang:17}. 
These multisoliton states can lose stability as the detuning is scanned, leading to a reduction in soliton number and a rearrangement of the remaining pulses~\cite{Gomila2022}. These transitions manifest experimentally as soliton steps—discrete drops in intracavity optical power~\cite{Herr_2014,Kippenberg_2018,Pasquazi_2018}. These observations suggest that the loss of crystal stability reflects changes in the interactions among its constituent solitons. Indeed, varying the detuning modifies the soliton profiles and, consequently, their overlap and mutual interactions.

The connection between pulse interactions and multipulse stability has been exploited in parabolic evolution equations and Hamiltonian systems~\cite{Elphick1990,Sandstede1998,Kapitula2004,Coles2010}, but only in the forward direction: given knowledge of the interaction, the stability properties can be found.

In this Letter, we turn this connection into an inverse method: we reconstruct the pairwise interaction between DKSs directly from the linear stability spectrum of a PSC, which can be obtained using standard tools. We show that a PSC behaves as an overdamped lattice whose positional-mode eigenvalues encode the pairwise interaction potential. This allows the interaction potential to be reconstructed directly from linear stability analysis, without tracking pair dynamics or assuming a functional form for the interaction. This approach parallels inverse lattice dynamics in condensed-matter physics, where microscopic interactions are inferred from phonon or relaxation spectra~\cite{Mohr2007,Flensburg,Keim2004,Baumgartl2008}. In contrast, existing methods to derive soliton interactions~\cite{Parra-Rivas_soliton_interaction,Cai1993,Manton1979} require system-specific calculations that can become cumbersome for general dispersion and do not exploit the crystal's collective response.

Our method is very general and applies beyond  conventional quadratic-dispersion. It predicts a soft soliton crystal state, in which the crystal stiffness vanishes and the relaxation time diverges, making it particularly susceptible to noise. Finally, it provides a clear physical interpretation of the soliton steps discussed above: they arise from a sign reversal of the pairwise-force derivative.

\begin{figure}[t!]
\centering
\includegraphics[width=0.45\textwidth]{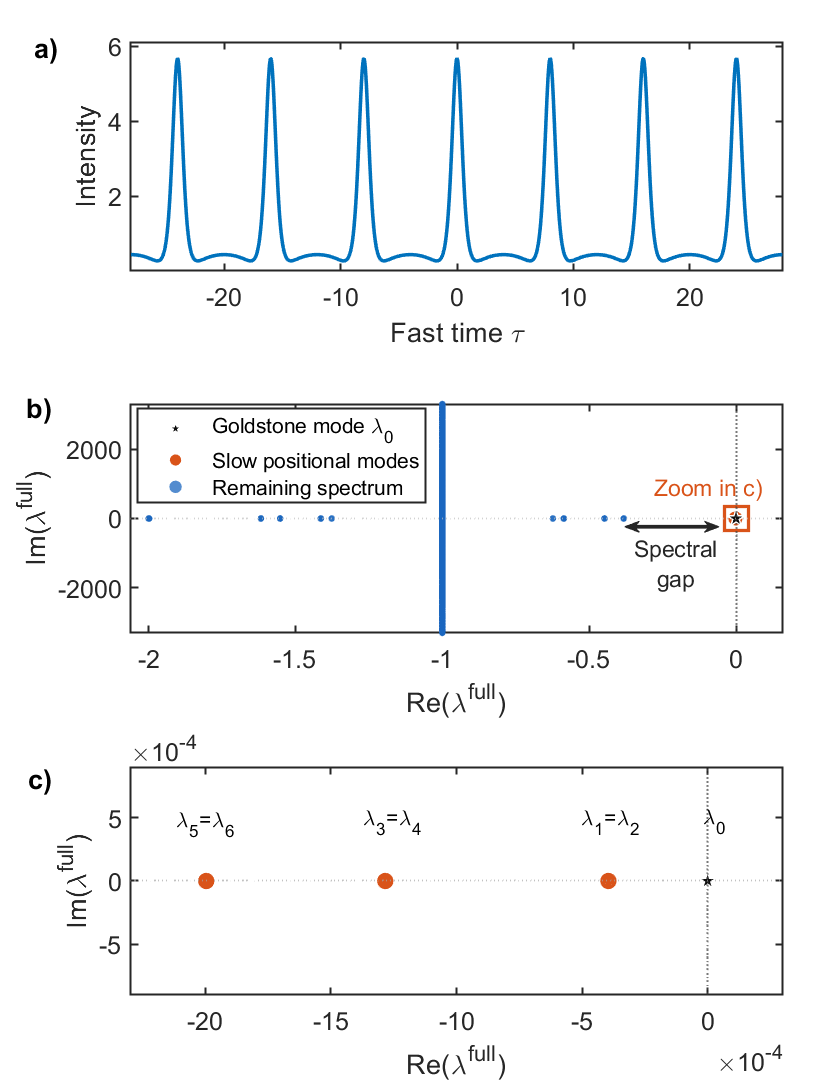}
\vskip -1mm
\caption{
(a) Intensity versus fast time for a stable PSC of seven solitons with nearest-neighbor separation $d=8$, for $\Delta=X=2.5$.
(b) Complete linear-stability spectrum of the crystal. The Goldstone mode, the six positional modes, and the remaining eigenvalues are indicated by a black star and orange and blue markers, respectively. The spectral gap between the positional eigenvalue with the smallest real part and the remaining eigenvalue with the largest real part is highlighted. The region enclosed by the orange box is enlarged in (c).
(c) Goldstone mode and six positional eigenvalues near the origin, forming three doubly degenerate real eigenvalues.
} 
\label{fig1}
\end{figure}
\textit{LLE and full stability analysis}--- We consider the intracavity field envelope $E(t,\tau)$ in a high-$Q$ ring microresonator, governed by the normalized LLE~\cite{LLE_1987,Coen2_2013,Pasquazi_2018}:
\begin{equation}
\frac{\partial E}{\partial t}=
\left[
-1+i\left(|E|^2-\Delta\right)
-i\eta\frac{\partial^2}{\partial\tau^2}
\right]E+\sqrt{X},
\label{eq:GLLE}
\end{equation}
where $t$ and $\tau$ denote the slow and fast times, respectively; $\Delta$ and $X$ are the normalized detuning and driving power; and $\eta=\operatorname{sgn}(\beta_2)$, with $\beta_2$ the second-order group-velocity-dispersion coefficient. We assume anomalous dispersion, for which $\eta=-1$.

Let $E_s(\tau)$ be a stationary solution of~\eqref{eq:GLLE}. To determine its linear stability, we write $E=E_s+\delta E$, where $\delta E=\delta E_{\mathrm{R}}+i\delta E_{\mathrm{I}}$ is a small complex perturbation, and linearize~\eqref{eq:GLLE} about $E_s$~\cite{Chembo}. Since the linearized dynamics couples the real and imaginary parts of $\delta E$, we introduce the vector perturbation $\delta\mathbf{E}=(\delta E_{\mathrm{R}},\delta E_{\mathrm{I}})^{\mathsf T}$. Seeking normal modes of the form $\delta\mathbf{E}=\mathbf{w}(\tau)e^{\lambda t}$ yields
\begin{equation}
\mathcal{J}_s\mathbf{w}=\lambda\mathbf{w},
\label{eq:full_LSA}
\end{equation}
where $\mathcal{J}_s$ is the Jacobian operator of the right-hand side of~\eqref{eq:GLLE}, evaluated at $E_s$, and $\lambda$ and $\mathbf{w}$ are the eigenvalues and eigenmodes, respectively. The solution $E_s$ is linearly stable if all non-neutral eigenvalues have negative real parts and unstable if at least one eigenvalue has a positive real part. We refer to this as the full linear stability analysis (full LSA) and hence denote the corresponding eigenvalues by $\lambda^{\mathrm{full}}$ in the following.

We consider a stationary PSC solution of the LLE~\eqref{eq:GLLE}, consisting here of seven equally spaced solitons (Fig.~\ref{fig1}(a)). The eigenvalue spectrum obtained from the full LSA is shown in Figs.~\ref{fig1}(b) and~\ref{fig1}(c). The neutral translational, or Goldstone, mode at the origin is the eigenvalue with the largest real part. Close to it lie six eigenvalues organized into three degenerate pairs, all with negative real parts, so the PSC is linearly stable (Fig.~\ref{fig1}(c)). An LLE stationary solution composed of $N$ well-separated solitons has $N$ eigenvalues close to the origin associated with the translations of its $N$ constituent solitons~\cite{Bengel}. One remains exactly at the origin owing to global translational invariance, while the positions of the other $N-1$ {\it positional eigenvalues}, are determined by intersoliton interactions~\cite{Bengel}. The rest of the spectrum lies much deeper in the left half-plane, producing the pronounced spectral gap highlighted in Fig.~\ref{fig1}(b). The magnitudes of the positional eigenvalues are two to three orders of magnitude smaller than that of the other eigenvalues (Figs~\ref{fig1}(b)-(c)). This indicates that positional shifts of the solitons relax much more slowly than all other perturbations, associated with the left part of the spectrum in Fig.~\ref{fig1}(b). Thus, the slow relaxation dynamics of the PSC is governed by its six non-neutral positional modes.

\begin{figure*}[t]
\centering
\includegraphics[width=1\textwidth]{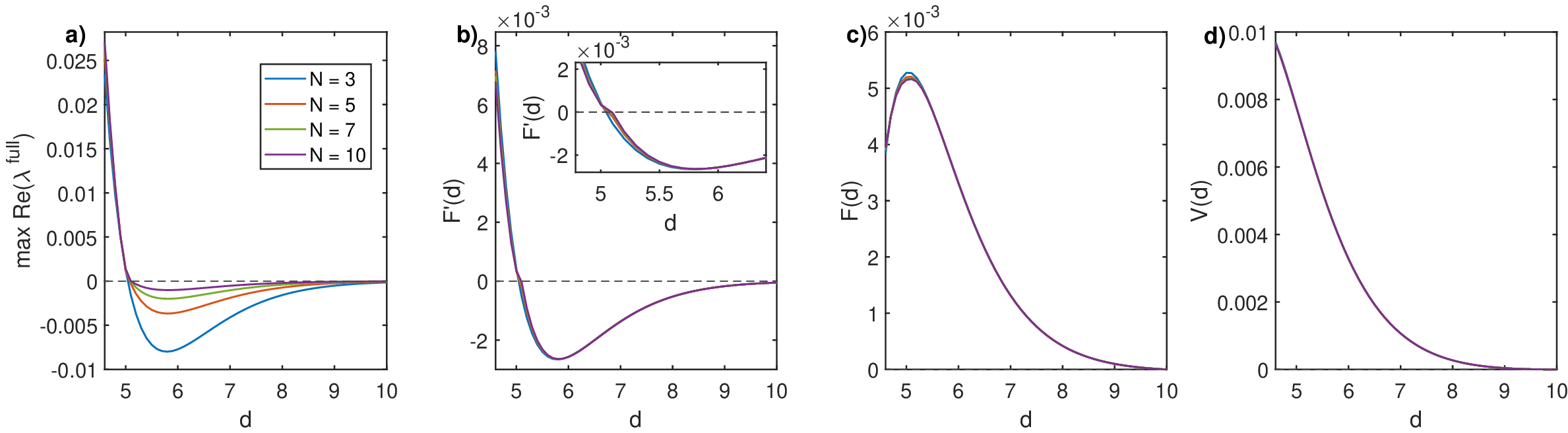}
\vskip -1mm
\caption{Reconstruction of the pairwise DKS interaction potential using PSCs with different soliton numbers $N$. (a) Largest real part of the full-LSA eigenvalues as a function of the intersoliton separation $d$. (b) Pairwise-force derivative reconstructed using~\eqref{eq:force_derivative}. (c) Reconstructed pairwise force. (d) Reconstructed interaction potential. All quantities in (b)--(d) are plotted as functions of $d$.
} 
\label{fig2}
\end{figure*}

\textit{Reduced LSA and reconstruction of the potential}---
We now consider only the slow positional modes and neglect the others.
We thereby adopt a reduced description restricted to soliton translations, modelling the PSC as a periodic chain of $N$ identical rigid solitons, equally spaced at equilibrium. Its only dynamical variables are the soliton center positions $\tau_j(t)$, with $j=1,\ldots,N$, measured along the fast-time coordinate $\tau$ and evolving on the slow time $t$. 
Let $\tau_{\mathrm{RT}}$ be the normalized round-trip time. The equilibrium spacing and soliton positions are then
\begin{equation}
d=\frac{\tau_{\mathrm{RT}}}{N},
\qquad
\tau_j^{(0)}=(j-1)d,
\qquad j=1,\ldots,N.
\end{equation}
We write $\tau_j(t)=\tau_j^{(0)}+\varepsilon_j(t)$, where $\varepsilon_j$ is the displacement of the $j$th soliton from its equilibrium position, with $|\varepsilon_j|\ll d$ and $\varepsilon_{j+N}=\varepsilon_j$. The nearest-neighbor separations are then
\begin{equation}
d_j=\tau_{j+1}-\tau_j
=d+\varepsilon_{j+1}-\varepsilon_j,
\end{equation}
where $\tau_{N+1}=\tau_1+\tau_{\mathrm{RT}}$.


For well-separated solitons, the interaction arises from the overlap of their exponentially decaying tails~\cite{Parra_Rivas_PRA2014}, allowing interactions beyond nearest neighbors to be neglected. The position of the $j$th soliton therefore evolves according to
\begin{equation}
\dot{\tau}_j
=
F(d_{j-1})-F(d_j),
\qquad j=1,\ldots,N.
\label{eq:N_soliton_dynamics}
\end{equation}
where $F(d)$ denotes the pairwise force between two DKSs separated by a distance $d$. The soliton dynamics is overdamped, as derived in the Supplemental Material by projecting the interaction-induced perturbation onto the adjoint translational mode~\cite{Leshem,Vladimirov2021}. Expanding the interaction force about the equilibrium separation, substituting the result into~\eqref{eq:N_soliton_dynamics}, and retaining only terms linear in $\varepsilon_j$ yields the equation of motion for the displacement of soliton $j$:
\begin{equation}
\dot{\varepsilon}_j
=
F'(d)
\left(
2\varepsilon_j-\varepsilon_{j-1}-\varepsilon_{j+1}
\right).
\label{eq:N_soliton_linearized}
\end{equation}
Defining $\boldsymbol{\varepsilon}=(\varepsilon_1,\ldots,\varepsilon_N)^{\mathsf T}$, the linearized dynamics becomes
\begin{equation}
\dot{\boldsymbol{\varepsilon}}
=
F'(d)\mathbf{M}_N\boldsymbol{\varepsilon},
\label{eq:N_soliton_matrix_dynamics}
\end{equation}
where $\mathbf{M}_N$ is the Laplacian matrix of the periodic chain, with elements $(\mathbf{M}_N)_{jk}=2\delta_{jk}-\delta_{j,k+1}-\delta_{j,k-1}$ and indices understood modulo $N$. Equation~(\ref{eq:N_soliton_matrix_dynamics}) describes the overdamped dynamics of the displacements of $N$ solitons about their equilibrium positions. This dynamics mirrors that of colloidal crystals: motion is overdamped, and short-range interactions yield a nearest-neighbor lattice description~\cite{PRR_colloids1}. Accordingly, the positional eigenvalues are purely real (Fig.~\ref{fig1}(b)) and correspond to relaxation rates rather than phonon frequencies~\cite{Keim2004,Baumgartl2008}.

The stability of the system described by \eqref{eq:N_soliton_matrix_dynamics} is determined by the eigenvalues of the dynamical matrix $F'(d)\mathbf{M}_N$. We refer to this analysis as the reduced LSA. The corresponding eigenvalues are~\cite{davis1994circulant}
\begin{equation}
\lambda_m^{\mathrm{red}}=F'(d)\gamma_m,
\qquad
\gamma_m=4\sin^2\left(\frac{\pi m}{N}\right),
\label{eq:reduced_eigenvalues}
\end{equation}
with $m=0,\ldots,N-1$ and $\gamma_m$ the eigenvalues of (circulant) matrix $\mathbf{M}_N$. Since $\gamma_m\geq0$, with $\gamma_0=0$ corresponding to the Goldstone mode, PSC stability is determined by the sign of the pairwise-force derivative $F'(d)$.

For well-separated solitons, for which the nearest-neighbor interaction holds, the full and reduced LSAs are equivalent and must yield the same dominant eigenvalue:
\begin{equation}
\max\operatorname{Re}\!\left[\lambda^{\mathrm{full}}(d)\right]
=
\max\lambda^{\mathrm{red}}(d),
\label{eq:LSA_matching}
\end{equation}
where the Goldstone mode is excluded from both spectra. Combining~\eqref{eq:reduced_eigenvalues} and~\eqref{eq:LSA_matching} yields
\begin{equation}
F'(d)
=
\frac{
\max\operatorname{Re}\!\left[\lambda^{\mathrm{full}}(d)\right]
}{
\bar{\gamma}
}.
\label{eq:force_derivative}
\end{equation}
Here, $\bar{\gamma}=\gamma_1$ for $F'(d)<0$, because all non-Goldstone $\lambda_m^{\mathrm{red}}$ are then negative and the dominant one, i.e., the one closest to the origin, corresponds to the smallest nonzero $\gamma_m$ (see \eqref{eq:reduced_eigenvalues}). For $F'(d)>0$, all non-Goldstone $\lambda_m^{\mathrm{red}}$ are positive, and the dominant one corresponds to the largest $\gamma_m$. Hence, $\bar{\gamma}=\gamma_{N/2}$ for even $N$ and $\bar{\gamma}=\gamma_{(N-1)/2}$ for odd $N$.

Equation~(\ref{eq:force_derivative}) is a reconstruction formula for the derivative of the pairwise interaction force between two solitons separated by $d$, using the LSA of a PSC with that spacing, independent of $N$. We can therefore fix $N$ and obtain the corresponding PSC stability curve, $\max \operatorname{Re}\lambda$ as a function of $d$. Then, $F'(d)$ is found by rescaling this curve by the corresponding $\bar{\gamma}$ values.

\begin{figure}[t!]
\centering
\includegraphics[width=0.45\textwidth]{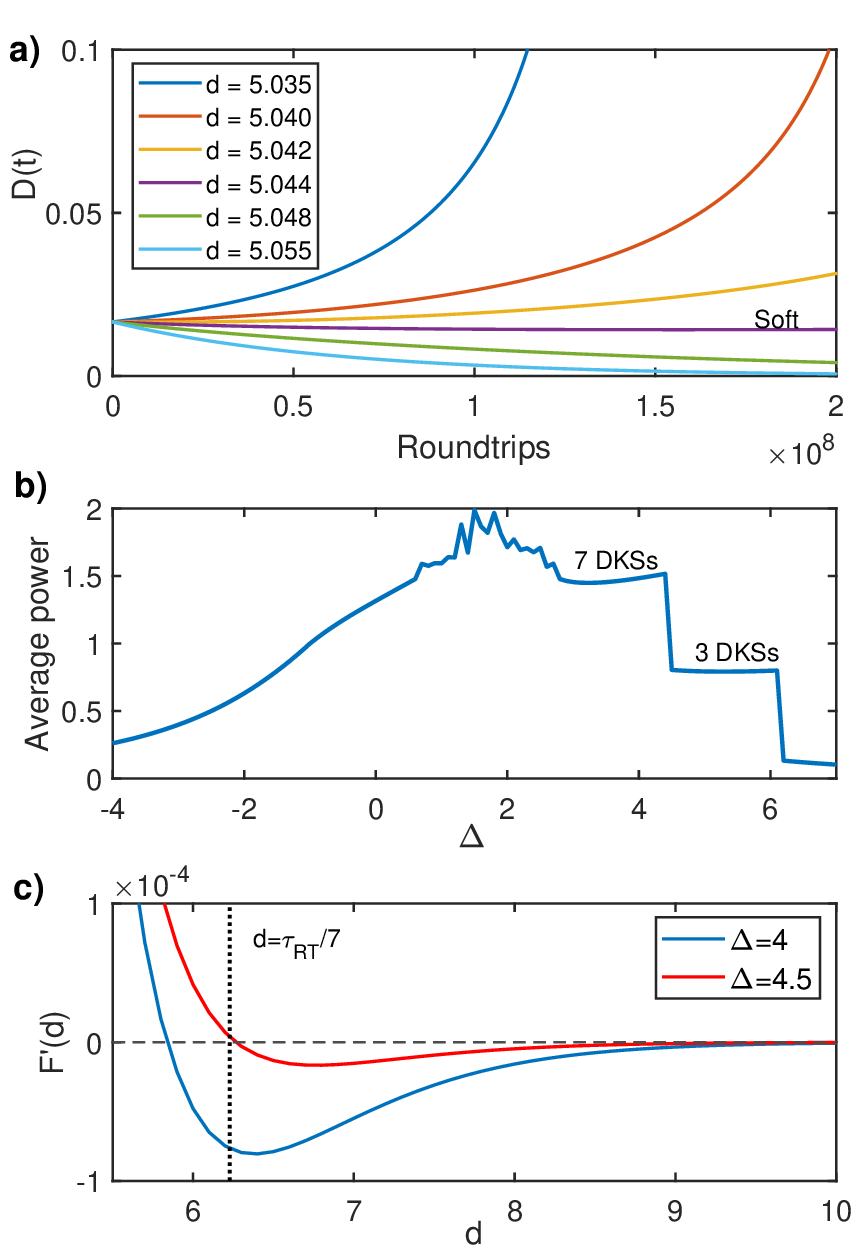}
\vskip -1mm
\caption{
Soft PSC and physical origin of soliton steps. (a) Evolution of the positional perturbation amplitude $D(t)$ for different intersoliton separations near the soft PSC. (b) Average intracavity power versus detuning, obtained by solving~\eqref{eq:GLLE} for $X=5$, showing a soliton step associated with a transition between states with different numbers of DKSs. (c) Reconstructed pairwise-force derivative for a seven-soliton PSC at $\Delta=4$ (blue curve) and $4.5$ (red), with $X=5$. The vertical dashed line indicates the seven-soliton spacing $d=\tau_{\mathrm{RT}}/7$.
} 
\label{fig3}
\end{figure}

Stability curves $\max\operatorname{Re}[\lambda^{\mathrm{full}}(d)]$ for different values of $N$ are shown in Fig.~\ref{fig2}(a). As $N$ increases, their negative portions approach the horizontal axis because they scale as $\gamma_1=4\sin^2(\pi/N)$, as verified at fixed $d$ in the Supplemental Material. By contrast, the positive parts scale as $\gamma_{N/2}=4$ for even $N$ and as $\gamma_{(N-1)/2}$ for odd $N$. The latter lies between $3$ and $4$ and approaches $4$ with increasing $N$. This causes the positive portions of the curves in Fig.~\ref{fig2}(a) to nearly overlap.
Rescaling each curve by the appropriate factor $\bar{\gamma}$ according to~\eqref{eq:force_derivative} collapses them onto the universal curve $F'(d)$, as shown in Fig.~\ref{fig2}(b). The curves are indistinguishable for $d\gtrsim5.8$, indicating that the collapse is accurate in this region. Small deviations arise at shorter separations (see inset in Fig.~\ref{fig2}(b)). As the solitons become more closely spaced, interactions with second and more distant neighbors, together with many-body corrections~\cite{Stone2013}, become increasingly relevant, reducing the validity of the nearest-neighbor description. This is consistent with the independent tests reported in Sec.~III of the Supplemental Material. These quantify the gradual breakdown of the reduced nearest-neighbor description as the soliton separation decreases. 
Since the width of an isolated soliton, measured between its first crossings of the background level, is approximately $3.1$ for these parameters, the model remains accurate even at separations smaller than twice the soliton width.

Once $F'(d)$ is known, we find the pairwise force by integration, imposing $F(\infty)=0$. This integration suppresses most of the small discrepancies among the curves, highlighting an advantage of reconstructing the force from its derivative, as shown in Fig.~\ref{fig2}(c). The interaction potential is then obtained from $F(d)=-V'(d)$, with $V(\infty)=0$. Here again, integration reduces residual discrepancies, leading to complete overlap among the curves and corroborating that the reconstruction is independent of $N$. We compared the reconstructed potential in Fig.~\ref{fig2}(d) with that obtained using the method of Ref.~\cite{Parra-Rivas_soliton_interaction}, finding excellent agreement for $d>4.5$ and deviations only at shorter separations, where the approximations underlying both methods become less accurate. 

\begin{figure}[t!]
\centering
\includegraphics[width=0.5\textwidth]{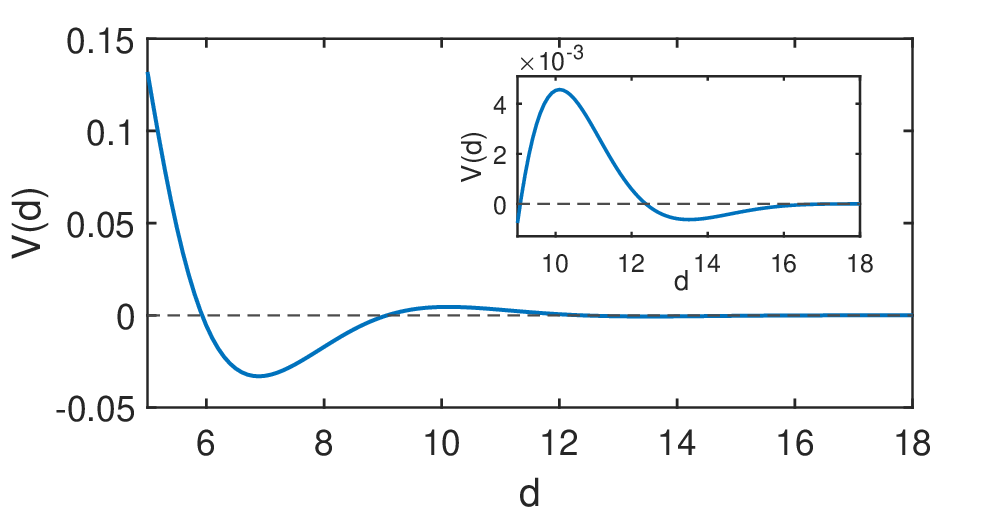}
\vskip -1mm
\caption{Reconstructed interaction potential for pure-quartic DKSs. The reconstruction is performed for $X=\Delta=2.5$ using a five-soliton PSC. The inset enlarges the region $9\leq d\leq18$, revealing additional oscillations of the potential.
} 
\label{fig4}
\end{figure}

Beyond reconstructing the interaction, our theory directly links the PSC stability curve to the pairwise force, allowing its features to be interpreted in terms of the microscopic interactions between the constituent solitons. For instance, the stability curves in Fig.~\ref{fig2}(a) cross zero at $d\approx5.04$, indicating marginal stability at this spacing. Our theory identifies this point with $F'(d)=0$, where the crystal stiffness vanishes according to~\eqref{eq:N_soliton_matrix_dynamics}. To clarify the physical meaning of this, we consider three-soliton PSCs with spacings near $d=5.04$ and apply a small nonuniform positional perturbation. We quantify the resulting deformation as $D(t)=\sqrt{\sum_{j=1}^{N}[d_j(t)-d]^2}$, where the $d_j(t)$ are the instantaneous separations between each soliton pair. $D(t)$ is invariant under rigid translations and, in the linear regime, evolves as $D(t)\propto\exp\{\max\operatorname{Re}[\lambda^{\mathrm{full}}(d)]t\}$. As shown in Fig.~\ref{fig3}(a), $D(t)$ decays to zero in the stable regime, recovering the equally spaced configuration; grows in the unstable regime, leading to soliton annihilation; and exhibits a plateau at the softening point $d=5.044$, where the relaxation rate vanishes. Thus, marginal stability corresponds to a vanishing stiffness and a diverging relaxation time, making the resulting soft PSC highly susceptible to noise.

Our theory also provides insight into soliton steps, a hallmark experimental signature of DKSs: discrete drops in intracavity power associated with a reduction in soliton number as the detuning is increased at fixed pump power~\cite{Pasquazi_2018,Karpov_2019,Herr_2014,HerrPRL2014}. Dynamical LLE simulations reproduce this behavior: for $X=5$, for instance, a seven-soliton PSC transitions to a three-DKS state near $\Delta=4.5$ (Fig.~\ref{fig3}(b)). Here, we use parameters representative of an $\mathrm{MgF}_2$ resonator with a free spectral range of $35.2~\mathrm{GHz}$ and $Q\approx4\times10^8$, which yield $\tau_{\mathrm{RT}}=43.6$~\cite{Herr_2014,Pasquazi_2018}. The full LSA confirms that the seven-soliton PSC loses stability near $\Delta=4.5$, in agreement with the dynamical simulation. According to~\eqref{eq:reduced_eigenvalues}, this loss of stability corresponds to a sign change of the pairwise-force derivative $F'(d)$. Indeed, reconstructing $F'(d)$ at $\Delta=4$, where the PSC is stable, and at the instability onset near $\Delta=4.5$ shows that, at the relevant lattice spacing $d=\tau_{\mathrm{RT}}/7$, $F'(d)$ is negative in the stable regime and changes sign at $\Delta=4.5$ (Fig.~\ref{fig3}(c)). Thus, the soliton step can be traced to change of the soliton profiles with detuning, which modifies their interaction and reverses the slope of the pairwise force. The transition is therefore deterministic, in the sense that its onset detuning can be quantitatively predicted. This instability can be further understood by considering one soliton in the PSC and its two nearest neighbors. At equilibrium, the forces exerted by the neighbors cancel, whereas a small displacement $\varepsilon_j$ produces a net force
$F(d+\varepsilon_j)-F(d-\varepsilon_j)\simeq2F'(d)\varepsilon_j$.
For $F'(d)<0$, this force is restoring; when $F'(d)>0$, it instead amplifies the displacement, driving the soliton toward one of its neighbors and triggering collisions and annihilations that reduce the soliton number.


Our reconstruction method requires no assumption about the functional form of the potential and can be extended beyond the quadratic dispersion case, provided that the nearest-neighbor approximation remains valid. As an example, we consider pure quartic dispersion~\cite{Blanco-Redondo_2016}. Pure-quartic DKSs, i.e., localized solutions of the quartic LLE~\cite{Taheri_2019}, have exponentially decaying, oscillating tails~\cite{ParraRivas_2022}. 
Applying the same procedure, the resulting potential, is shown in Fig.~\ref{fig4}. It differs qualitatively from its conventional quadratic-dispersion counterpart by exhibiting multiple oscillations, highlighted in the inset of Fig.~\ref{fig4}, inherited from the soliton tails~\cite{Taheri_2019,ParraRivas_2022,Panda26}. This resembles the oscillatory soliton tails and interaction potential of the standard quadratic-dispersion LLE at low detuning and pump power~\cite{Parra-Rivas_soliton_interaction,Cai1993}. The resulting sequence of alternating maxima and minima is consistent with the recently reported quantization of soliton-pair separations, corresponding to alternating stable and unstable pair solutions~\cite{Hetzel26}. We verified using the indicators presented in Sec.~III of the Supplementary Materials that the reconstruction achieves an accuracy comparable to that found for quadratic dispersion and exhibits a similar dependence on soliton separation, confirming that the method applies equally well to pure-quartic dispersion.

\textit{Conclusion}---We have established a direct connection between the linear stability spectrum of soliton crystals and the pairwise interactions among their constituent DKSs. Mapping the slow positional modes onto those of an overdamped periodic lattice enables the interaction force and potential to be reconstructed without assuming their functional forms.  Beyond reconstruction, the framework identifies a marginally stable PSC as a soft-crystal state with vanishing stiffness and reveals that soliton steps---widely observed experimental signatures of DKSs---are triggered by a sign reversal of the pairwise-force derivative. Applying the framework to pure-quartic DKSs demonstrates its ability to capture qualitatively different interactions induced by nonstandard dispersion. 
More broadly, our results establish linear stability analysis as an inverse probe of microscopic interactions in dissipative soliton ensembles, providing a conceptual bridge to inverse lattice dynamics in condensed-matter physics. Our numerical approach opens a route to studying nonlinear-wave interactions in systems where existing analytical methods are difficult to apply, including fractional-dispersion solitons~\cite{hoang2025nonlinear,liu2023experimental,Ha_2026}, Fabry--Pérot microresonators~\cite{Wildi23}, media with quadratic $\chi^{(2)}$ nonlinearities~\cite{Bruch2021,Roy2022}, and active cavities~\cite{Columbo2021,Kazakov2025}.\\\\
\textit{Acknowledgments} --- The authors thank Prof.~David 
McKenzie and Dr~Umbertoluca Ranieri for fruitful discussions.\\
\textit{Funding} ---
Australian Research Council (CE230100006, DE220100509, DP230102200); Royal Society Te Apārangi Marsden fund 23-UOA-053.

\bibliography{Soliton_inter}
\end{document}